\documentclass[twocolumn,prl,showpacs,preprintnumbers,amsmath,amssymb,superscriptaddress]{revtex4}
\usepackage{braket}
\usepackage{graphicx,color}
\usepackage{amsmath}

\newcommand\bx{{\bf x}}

\newcommand\rF{{\rm F}}

\newcommand{\lk}{\left(}
\newcommand{\rk}{\right)}

\newcommand{\ldk}{\left[}

\newcommand{\rdk}{\right]}
\newcommand\beq{ \begin{eqnarray} }
\newcommand\eeq{ \end{eqnarray} }

\usepackage{ulem}

\begin{document}
\title{
Extracting non-local inter-polaron interactions from collisional dynamics
}
\author{Junichi Takahashi}
\affiliation{Department of Electronic and Physical Systems, Waseda University, Tokyo 169-8555, Japan}

\author{Hiroyuki Tajima}
\affiliation{Department of Mathematics and Physics, Kochi University, Kochi 780-8520, Japan}
\affiliation{Quantum Hadron Physics Laboratory, RIKEN Nishina Center, Wako, Saitama 351-0198, Japan}

\author{Eiji Nakano}
\affiliation{Department of Mathematics and Physics, Kochi University, Kochi 780-8520, Japan}

\author{Kei Iida}
\affiliation{Department of Mathematics and Physics, Kochi University, Kochi 780-8520, Japan}

\date{\today}

\begin{abstract}
This study develops a novel experimental method of deducing the profile
of interaction induced between impurities in a trapped gas of ultracold
Fermi/Bose atoms, which are often referred to as Fermi/Bose polarons.
In this method, we consider a two-body Fermi/Bose polaron collision
experiment in which impurities and atoms interact only weakly.
Numerical simulations of the quantum dynamics reveal the possibility
to obtain information regarding the non-local induced interaction between
two polarons from a measured profile of the polaron wave packet
at several snapshots.  This is because the potential of the induced
interaction is well balanced by the quantum potential whenever the WKB
approximation for the relevant Schr\"odinger equation is applicable.
\end{abstract}


\maketitle


{\it Introduction}---
To describe interparticle interactions is indispensable in various fields of physics.
At extremely high energy, fundamental interactions are mediated by gauge bosons \cite{Book_Weinberg1995}, while in atomic nuclei, the nuclear force at a distance occurs through exchange of mesons \cite{Taketani1951}.
In conventional superconductors,
attractive electron--electron interactions are induced by phonons \cite{Bardeen1957}.
In high-$T_{\rm c}$ superconductors, furthermore, background spin and/or charge fluctuations are expected to play a role \cite{Scalapino1995}.

Ultracold atomic systems can be used as a platform for studying medium-induced interactions. Particularly, such systems can exhibit high tunability and pureness, thereby serving as excellent quantum simulators for studying quantum many-body theories \cite{Bloch2008, Giorgini2008, Dalfovo1999}. For example, cold atomic Fermi systems with a long scattering length close to the unitary limit are good simulators with regard to low density neutron matter \cite{Horikoshi2019}. They also possess the unique feature of having the real-time dynamics and momentum distribution measured \cite{Pethick-Smith}.
Moreover, atomic Fermi/Bose polarons, which are quasiparticles that consist of impurities (minority atoms) immersed in a degenerate Fermi gas \cite{Schirotzek2009, Frohlich2011, Kohstall2012} or a Bose-Einstein condensate \cite{Jorgensen2016, Hu2016, Rentrop2016}, have been realized.
Since the impurities have the mass and the interaction with each other
modified from those in vacuum by the impurity-medium interaction, a
great deal of research has been conducted both from theoretical and experimental sides.
Theoretically, the Ruderman-Kittel-Kasuya-Yoshida (RKKY) type \cite{Ruderman1, Kasuya1, Yoshida1} and Yukawa/Efimov type interactions have been proposed as typical examples of non-local induced interactions between
%
heavy Fermi and Bose polarons \cite{Nishida2009, Suchet2017, Heiselberg2000, Yu2012, Enss2020, Naidon2018}, respectively.
In connection with such induced interactions, moreover, bi-polaron formation for Bose polarons \cite{Naidon2018, Camacho-Guardian2018, Dehkharghani2018, Volosniev2015, Mistakidis2019} and the energy shift and broadening of Fermi polarons \cite{Giraud2012, Mistakidis2019c, Tajima2018} have been actively discussed.
Recently, the presence of non-local interactions between Fermi polarons has been experimentally confirmed \cite{DeSalvo2019, Edri2020}.
Here, the formation of solitons through the corruption of a Bose gas was found to be different from the behavior predicted when only local interactions occur in a boson-fermion mixture, which is an evidence for the existence of a non-local interaction mediated by fermions.
The non-local nature of such interactions, however, remains to be investigated experimentally.   Given that similar induced interactions occur in various fields of physics, therefore, real cold atom experiments in this direction would serve as a cornerstone.

We now address the question ``What is enough information to find induced interactions between particles in a medium?''
For concreteness, we consider the case in which two polarons are approximately described by a two-body time-dependent Schr\"odinger equation (TDSE)
with a mediated interaction instead of dealing with the medium explicitly.
In this case, the simplest answer is a two-body polaron wavefunction.
That is, the interaction can be obtained from the wavefunction by transforming the TDSE as
$
 U(r)
 = \Psi^{-1}
    \ldk i\hbar \frac{\partial}{\partial t}
          - h_1(\bx_1) - h_2(\bx_2) \rdk \Psi
$,
where
$\Psi$
is the wavefunction,
$h_i$ denotes the one-particle Hamiltonian,
$r=|\bx_1 - \bx_2|$,
and
$U$ denotes the interaction between two polarons.
This means that one can in principle constitute
the inter-polaron interaction
$U$ as function of the relative coordinate if the $\Psi$ keeps nonvanishing
in the range of
$U(r)\not=0$
throughout such
experiments as the collisional ones \cite{Sommer2011, Joseph2011, Valtolina2017, Reynold2020}.
In reality, however, it is difficult to simultaneously
obtain the phase and amplitude of the wavefunction by experiments.
In fact, at a given snapshot, only the amplitude of the wavefunction
can be measured via the square root of the probability
density of the wavefunction.  For trapped cold atomic systems,
one can use, e.g., a wave packet of a non-interacting Bose gas
as an impurity and then identify the number density of
the Bose gas with the probability density of the
impurity.  In this study, we propose a method of deducing
the non-local interaction between two such impurities
in a medium from the density of the Bose gas at several snapshots.
It is noteworthy that the idea of deriving bulk thermodynamic
quantities from the density of a single-species cold atomic gas
has been already proposed \cite{Ho2010}. Despite similarity
in strategy, there is a crucial difference from our proposal.
Authors in Ref.~\cite{Ho2010} use the density distribution
to obtain the thermodynamic properties of the corresponding
homogeneous system within local density approximation (LDA),
while information about the non-local interaction between
atoms in a trap, which is of interest here, is out of the reach of LDA.

{\it Formulation}---
For future possible experimental realizations, we focus on a two-polaron collision in a harmonic trap filled with a Fermi/Bose atomic gas.
Each impurity, which is a Bose condensate wave packet,
is initially localized by a separate confining harmonic potential,
released at $t=0$, and allowed to collide with another impurity.
Note that experimental setups that involve a scenario highly similar to that considered
in this study have been proposed \cite{Mistakidis2019b, Magierski2019, Kwasniok2020, Tajima2020},
while a part of them have been realized in experiment by using optical tweezers \cite{Reynold2020}.


We assume that at $t = 0$, each impurity has already been immersed in the majority
gas long enough to form a polaron and that at $ t \geq 0$ a medium-induced
interaction occurs between the polarons.  Furthermore, we consider the polarons to be
distinguishable particles that do not interact directly with each other and to remain
robust during the time evolution owing to coupling between the impurity and the majority gas
that is too weak for the entire system to be relaxed.  In a regime of strong coupling,
on the other hand, a hydrodynamic description looks more relevant.  In fact, a hydrodynamic
analysis of collisions of two polaronic clouds in a similar setup has been performed
\cite{Tajima2020}; in this case, induced inter-polaron interactions have been incorporated
even inside a cloud but only within the LDA.


In the weak coupling regime of interest here,
the dynamics of the two polarons essentially obeys the TDSE: $i\hbar\partial_t \Psi = [h(\bx_1) + h(\bx_2) + U_{\rm med}(r)] \Psi$, where $h(\bx_i) = -\hbar^2\nabla^2_{\bx_i}/2m + V_{\rm HO}(\bx_i)$ with the trap potential $V_{\rm HO}(\bx_i)=m\omega^2 \bx_i^2/2$
and $U_{\rm med}(r)$ denotes an interaction mediated by the majority gas.
Here, for simplicity, we assume that the trap frequency $\omega$ and the bare mass $m$ of an impurity atom are the same between the polarons and that the dynamics is 1D.
We next rewrite the TDSE in the center-of-mass frame by using the relative coordinate $x=x_1-x_2$ as
$
i\hbar\frac{\partial\varphi}{\partial t}
= \ldk -\frac{\hbar^2}{2m_{\rm r}} \frac{\partial^2}{\partial x^2}
  + V_{\rm r, HO}(x) + U_{\rm med}(|x|) \rdk \varphi
$
where $m_{\rm r}=m/2$ is the reduced mass of the impurities, $V_{\rm r, HO}(x)=\frac{1}{2}m_{\rm r}\omega^2 x^2$,
and $\varphi(x, t)$ is the relative wavefunction of the polarons.
We then substitute the wavefunction $\psi=\sqrt{\rho(x,t)}\exp[iS(x,t)/\hbar]$
into the TDSE,
thereby obtaining the coupled equations that are the equation of continuity and the
quantum Hamilton--Jacobi equation \cite{Book_Tannor2007}
\beq \label{eq:QHJ}
C(S; x, t) + Q(\rho; x, t) + U_{\rm med}(|x|) = 0,
\eeq
with $C(S;x,t) = \frac{\partial S}{\partial t} + \frac{1}{2m_{\rm r}} \lk \frac{\partial S}{\partial x} \rk^2 + V_{\rm r, HO}$ and $Q(\rho: x,t)=-\frac{\hbar^2}{2m_{\rm r} \sqrt{\rho}} \frac{\partial^2 \sqrt{\rho}}{\partial x^2}$.
Here,
$C$ is the potential given by the phase fluctuations, while
$Q$ is the potential given by the density fluctuaions,
which is often referred to as the quantum potential.
Equation (\ref{eq:QHJ}) shows that the interaction
$U_{\rm med}$ is determined by the balance of
those potentials.

%

{\it Analysis and Results }---
We proceed to solve the TDSE for a particular interaction and
obtain density profiles of a polaron to calculate the quantum potential.
Since each polaron is initially trapped in the respective confining
harmonic potential, we set the initial two-polaron wavefunction
as a pair of the Gaussian wave packets via
$	\Psi(x_1, x_2, t=0) = \phi_+(x_1)\phi_-(x_2) $,
with
$
	\phi_{\pm}(x) = \frac{1}{(2\pi\eta^2)^{\frac{1}{4}}}
	\exp\ldk -\frac{(x \pm x_0)^2}{4\eta^2} \rdk,
$
where the two parameters $x_0$ and $\eta$ controls the initial position
and the width of the respective wave packet.
For such nodeless wave packets, the WKB approximation is expected to be good.
We solve the TDSE
under the corresponding initial condition
$
\varphi(x, t=0) = \frac{1}{(4\pi \eta^2)^{\frac{1}{4}}}
			   \exp\ldk -\frac{(x-2x_0)^2}{8\eta^2} \rdk,
$
by using the second-order split-step Fourier method.
We adopt (i) RKKY, (ii) Yukawa, and (iii) Efimov type interactions as $U_{\rm med}$.
We note that the interactions (i)--(iii), which are derived for a 3D
homogeneous medium, are used for simplicity and that the concrete form of the
interactions is not important for the present purpose of finding a
way of capturing the mediated interactions through experiments
unless the range of the interactions is significantly long.

\begin{figure}[ht]
\begin{center}
\includegraphics[width=8.5cm]{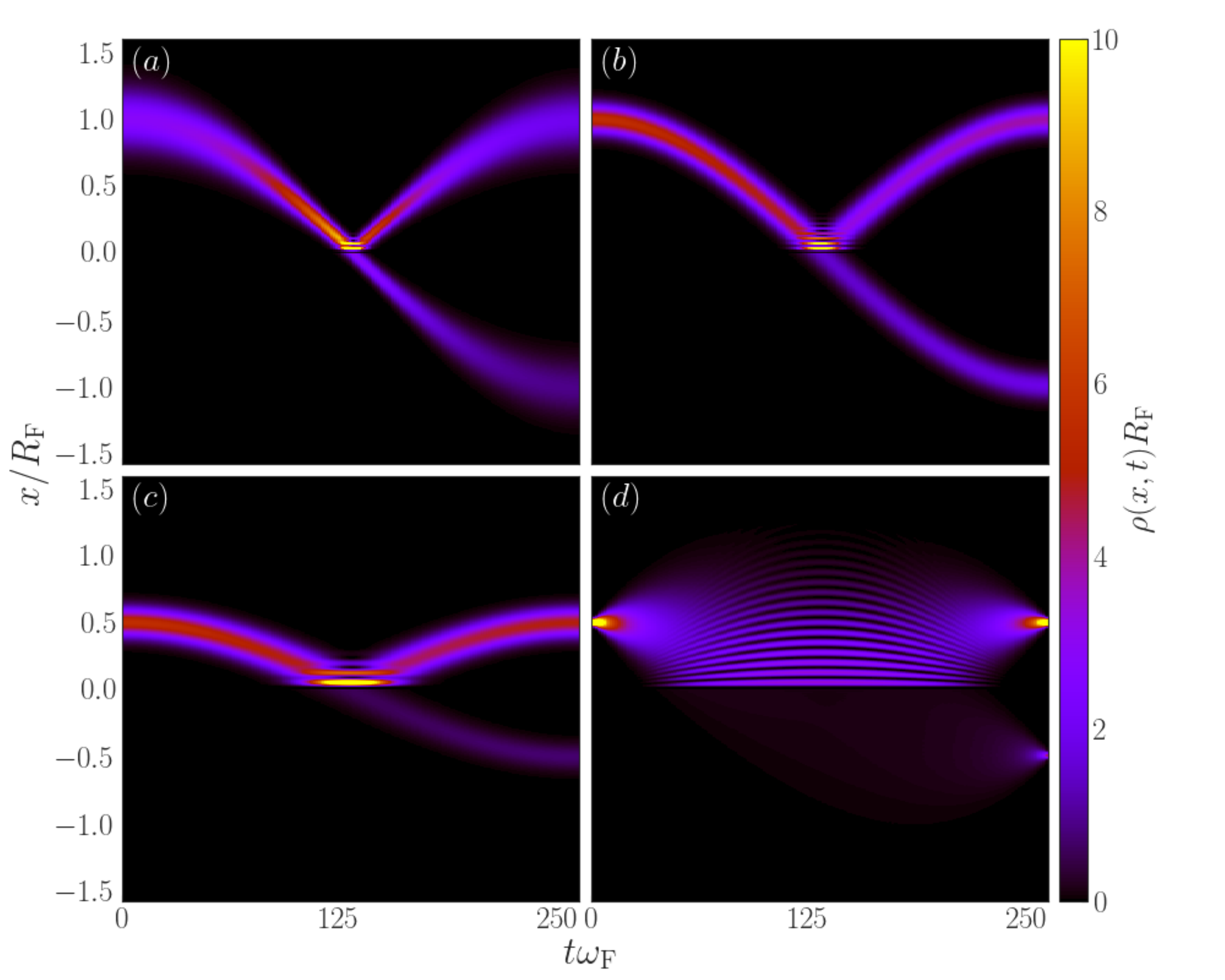}
\caption{
Probability density
of the two-polaron relative wave packet (time vs.\ relative position) for (a) $x_0/R_\rF=0.5, \, \eta/R_\rF=0.1$, (b) $x_0/R_\rF=0.5, \, \eta/R_\rF=0.05$, (c) $x_0/R_\rF=0.25, \, \eta/R_\rF=0.05$, and (d) $x_0/R_\rF=0.25, \, \eta/R_\rF=0.01$.
Here, $R_\rF$ and $E_\rF$ denote the Thomas--Fermi radius and Fermi energy, respectively.
The Fermi energy is given by
$E_\rF = \frac{1}{2} m \omega^2 R_\rF^2 = (6N)^{\frac{1}{3}}\hbar\omega$.
The time $t_c$ at which the center of the relative wave packet reaches $x=0$ can be estimated from
$
 t_c \omega_\rF
 \simeq \frac{1}{4} \frac{2\pi}{\omega} \omega_\rF
 \simeq 125,
$
where $\omega_\rF = E_\rF/\hbar$.
}
\label{fig:collisionaldynamics}
\end{center}
\end{figure}

(i) The RKKY type interaction is as follows:
\beq \label{eq:RKKY}
U_{\rm med}(|x|)
=\frac{\hbar^2 M a^2}{2 \pi M_{\rm r}^{2}}\frac{2k_{\rF} |x| \cos(2k_\rF |x|)-\sin(2k_\rF |x|)}{|x|^4},
\eeq
where $a$ is the $s$-wave scattering length between a majority atom of mass $M$ and an impurity, taken to be independent of which impurity the majority atom interacts with, $k_\rF$ is the Fermi wavenumber of the majority gas, and $M_{\rm r}=mM/(m+M)$.
Figure \ref{fig:collisionaldynamics} presents the probability
densities
for the relative two-polaron
wave packet before and after the collision of a pair of the polaron wave packets that
evolve from various initial conditions under the influence of $U_{\rm med}$.
%
We set the number of atoms in the majority gas as a typical one, $N=10^5$, while assuming $(k_\rF a)^2= 2$ and $M=m$.
It is interesting to note that before the collision,
the width of the two-polaron relative wave packet increases (decreases) with $t$ in the case of Fig.\ 1(d) (Fig.\ 1(a)),
while remaining almost unchanged with $t$ in the cases of Figs.\ 1(b) and 1(c).  This behavior stems from
the fact that the Gaussian wave packet for a free particle with mass $m_{\rm r}$ has the width increased
with time like $\sqrt2\eta\sqrt{1+(\hbar t/4m_{\rm r}\eta^2)^2}$,
while, as two polarons approach each other,
the confining potential acts to reduce the width of
the wave packet like $\hbar/\sqrt{2m_{\rm r}(E-V_{\rm r, HO})}$ with the relative energy $E$.
Incidentally, the collision time is of order $t_c$ (quarter of the dipole oscillation period) in the cases of Figs.\ 1(a)--(c),
while being well before $t_c$ in the case of Fig.\ 1(d).


\begin{figure}[ht]
\begin{center}
\includegraphics[width=8.5cm]{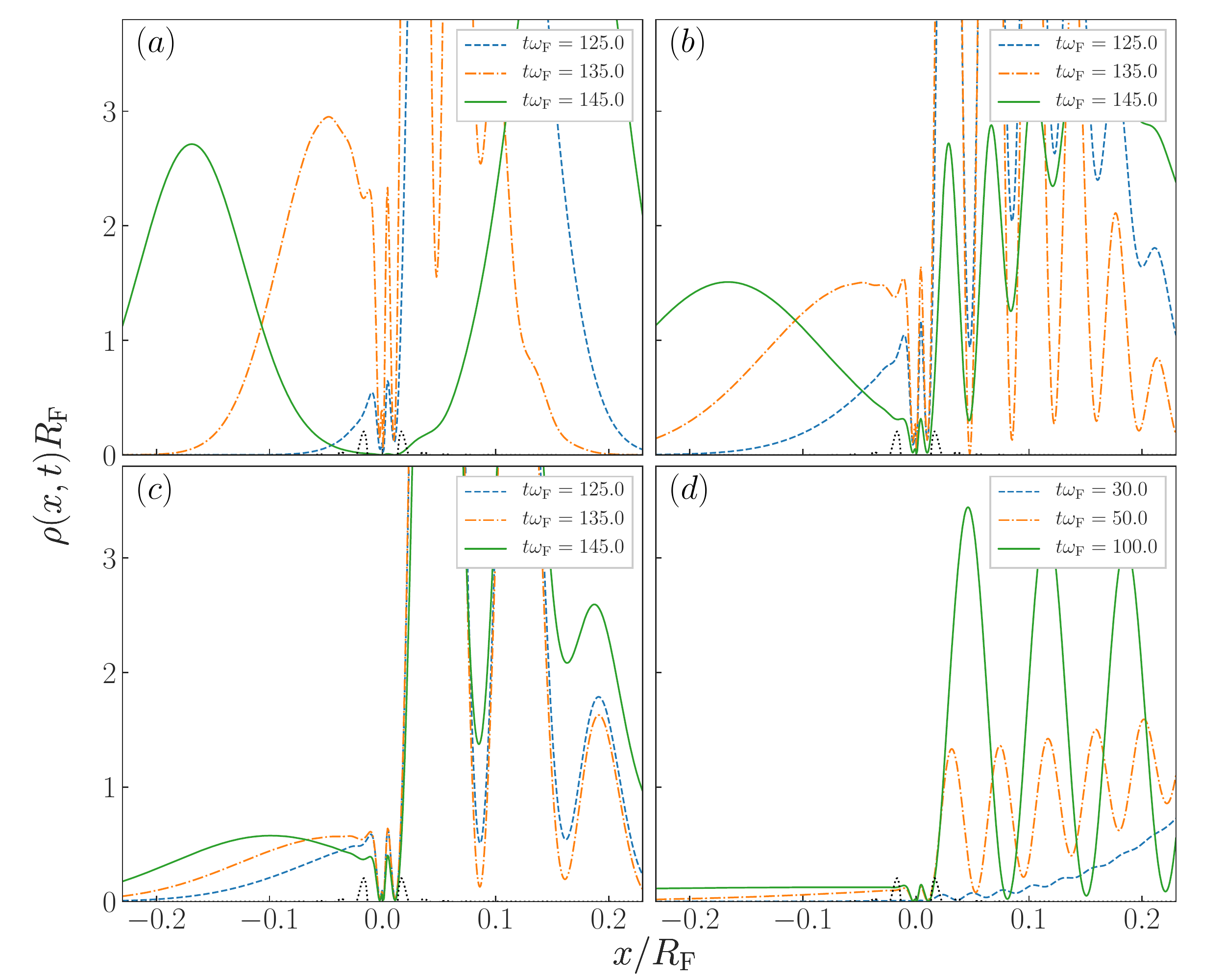}
\caption{
    Relative density profile $\rho(x,t)$ at three values of $t \omega_\rF$ of order or greater than the collision time.
    The distinction between (a)--(d) corresponds to the difference in initial parameters as in Fig.~\ref{fig:collisionaldynamics}.
}
\label{fig:density_2d}
\end{center}
\end{figure}

\begin{figure}[ht]
\begin{center}
\includegraphics[width=8.5cm]{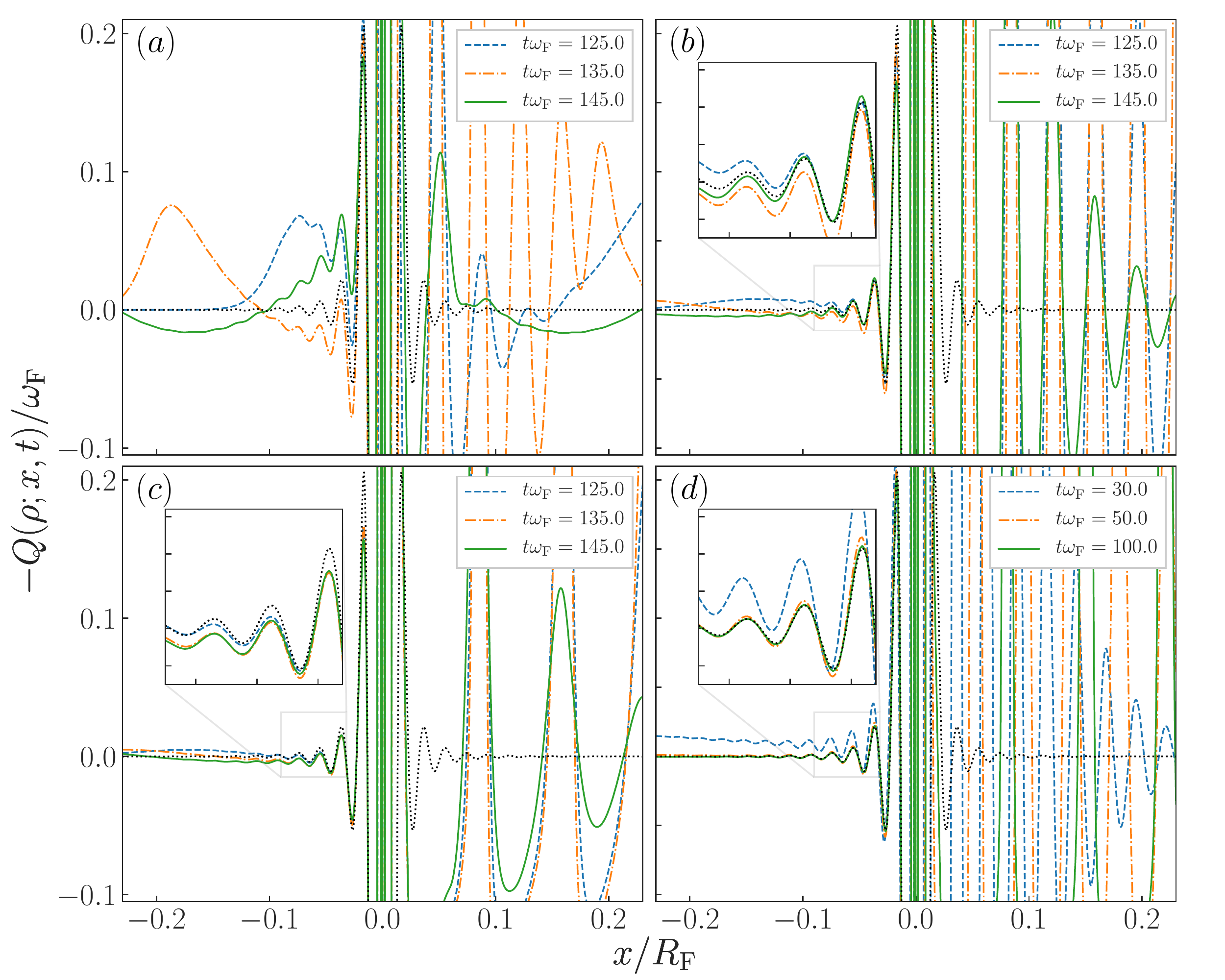}
\caption{
    Same as Fig.~\ref{fig:density_2d} for minus the quantum potential.
    For comparison, the RKKY interaction employed to solve the TDSE is also plotted in dotted line.
    For detailed movies, see~\cite{Animation}.
}
\label{fig:quantum_potential}
\end{center}
\end{figure}

Figure \ref{fig:density_2d} depicts the relative density
$\rho(x,t)$
at three different times, most of which are close to the collision time extracted from Fig.~\ref{fig:collisionaldynamics}.
Numerical data in the regions of $x<0$ and $x>0$ represent the behavior of the transmitted component and of the incident and reflected components of the relative wave packet, respectively.
Once this kind of profiles are obtaind by
experiments, one can use them
as input data in deriving the quantum potential.

We proceed to exhibit, in Fig.\ \ref{fig:quantum_potential}, minus the quantum potential $-Q(\rho; x,t)$ that can be derived from pseudo data for $\rho(x,t)$ shown in Fig.~\ref{fig:density_2d}.
One can observe from Fig.\ \ref{fig:quantum_potential}
that the quantum potential in the transmitted wave regime ($x<0$) well reproduces
the oscillating pattern of the RKKY interaction for any initial condition,
while the reproducibility of the RKKY interaction itself depends on the initial condition.
Comparing Figs.~\ref{fig:quantum_potential} (a)--(c), in which cases the width of the transmitted relative
wave packet just after the collision is different, we find it advantageous for the transmitted
wave packet to be sufficiently spread.
Remarkably,
Fig.~\ref{fig:quantum_potential} (d) shows an even better reproducibility of the interaction
for the initial condition that allows the two respective wave packets to spread and merge well before the centers of these packets come together at the origin at a timescale $t_c$ of the dipole oscillation.
This supports the tendency that the wider the transmitted wave packet, the higher the reproducibility of the interaction from the quantum potential.
This tendency in turn ensures $C(S;x,t)\simeq0$ and
$U_{\rm med}\simeq -Q(\rho; x,t)$ in the region of $x<0$ where the WKB approximation is a good approximation because of the sufficiently spread wave packet and gentle spatial dependence of the confining potential.

In the region of $x>0$ in each panel of Figs.\ \ref{fig:collisionaldynamics} and \ref{fig:density_2d}, on the other hand, a density oscillation pattern emerges after the collision.
This pattern stems naturally from interference between the incident and reflected components of the relative wave packet, while the presence of the reflected component arises
mainly
from a repulsive part of $U_{\rm med}$.
As can be seen from Fig.\ \ref{fig:quantum_potential}, however, the periodicity of this density oscillation
is totally different from that of $U_{\rm med}$.  In other words, such quantum interference leads inevitably to $C\not\simeq0$.  Then, the WKB approximation is no longer valid, and hence the mediated interaction cannot be reproduced by $-Q$.

\begin{figure}[ht]
\begin{center}
\includegraphics[width=8.5cm]{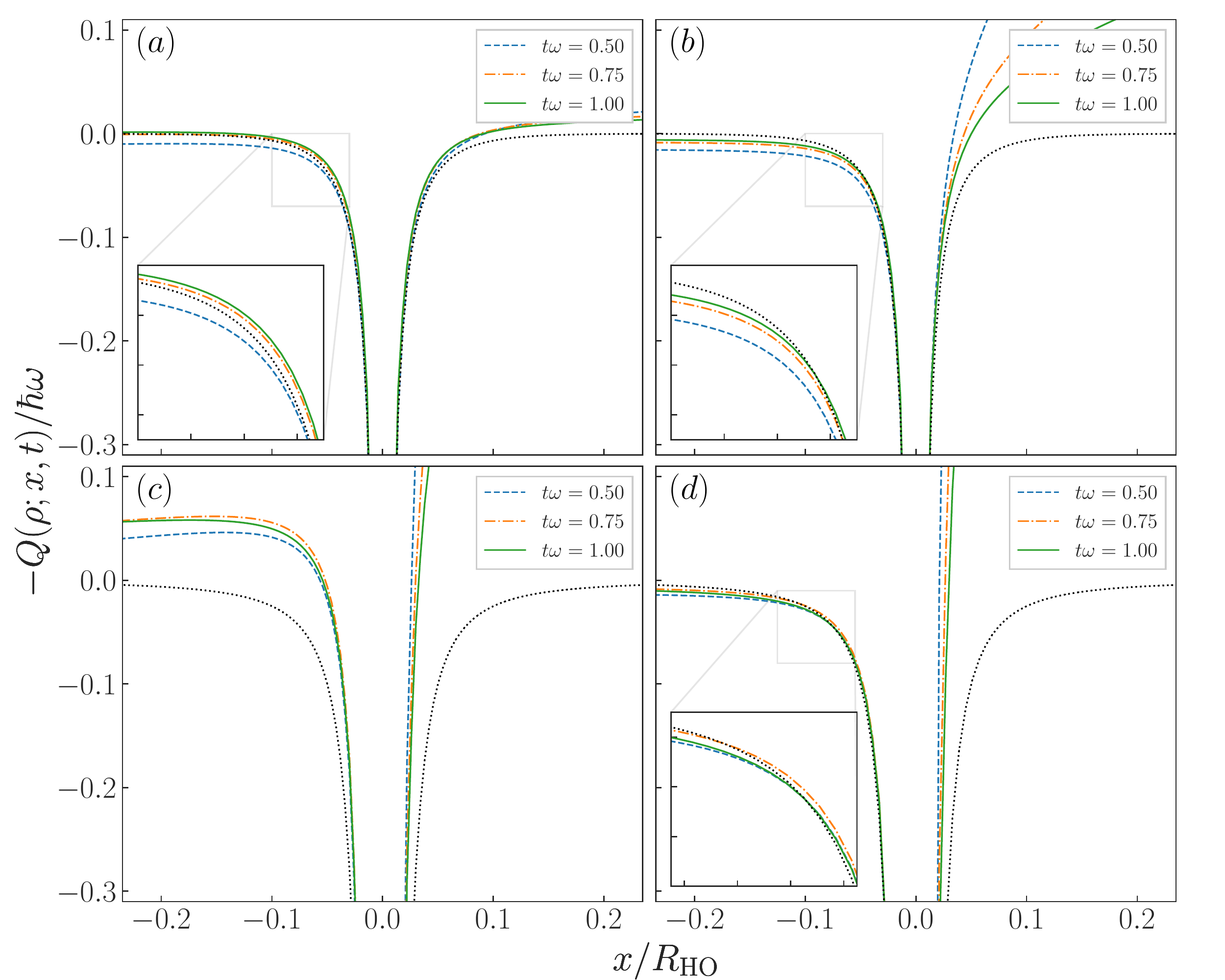}
\caption{
Minus the quantum potential at three different times as plotted for the Yukawa (upper panels) and Efimov (lower panels) type interactions
with $U_0/\hbar\omega=U_1/\hbar\omega=-0.1$ and $\kappa_0 R_{\rm HO}=\kappa_1 R_{\rm HO}=20$, where $R_{\rm HO}=\sqrt{\hbar/m_r \omega}$.
In panels (a) and (c) [(b) and (d)], the initial parameters are set to $x_0/R_{\rm HO}=0.25$ and $\eta/R_{\rm HO}=0.05$
($x_0/R_{\rm HO}=1$ and $\eta/R_{\rm HO}=0.05$).
The dotted line in each panel denotes the mediated interaction involved.
For detailed movies, see~\cite{Animation}.
}
\label{fig:quantum_potential_for_BP}
\end{center}
\end{figure}

Let us turn to the (ii) Yukawa and (iii) Efimov type interactions:
\beq
 U_{\rm med}(|x|) =
 \begin{cases}
   U_0 \frac{\exp(-\kappa_0 |x|)}{\kappa_0|x|} & ({\rm Yukawa} \, {\rm type}),\\
   U_1 \frac{\kappa_1^{-2}}{|x|^2}             & ({\rm Efimov} \, {\rm type}).
\end{cases}
\eeq
We again confine ourselves to sufficiently spread relative wave packets for the WKB
approximation to hold at least in the region of $x<0$.
Figure~\ref{fig:quantum_potential_for_BP} shows minus the relative quantum potentials
obtained by numerically simulating collision dynamics of two Bose polarons that
interact with (ii) or (iii).
We can observe that the quantum potential tends to well reproduce the respective mediated interaction in the
regime ($x<0$), just like the RKKY case shown in Fig.~\ref{fig:quantum_potential}.
Comparing Figs.~\ref{fig:quantum_potential_for_BP} (a) and (b), in which cases
the initial position of the relative wave packet is different under the same Yukawa interaction,
one can see that $ U_{\rm med} $ is more reproducible in the former than in the latter.
On the other hand, comparing Figs.~\ref{fig:quantum_potential_for_BP} (c) and (d),
which are the Efimov counterparts to Figs.~\ref{fig:quantum_potential_for_BP} (a) and (b),
one can realize that $ U_{\rm med} $ is more reproducible in the latter than in the former.
These results imply that tuning of the initial energy would be desirable for
better reproduction of $ U_{\rm med} $ in a manner that depends on the range of the mediated interaction.
We remark in passing that in contrast to the RKKY case,
almost
no density oscillation emerges
in the region of $x>0$ for the Yukawa and Efimov type interactions, whose purely attractive nature produces
almost
no reflected wave packet.  Then, the WKB approximation is expected to be valid
even for $x>0$.  A significant deviation of $-Q$ from $ U_{\rm med} $ that can be
observed in this region for the Efimov type interaction, therefore, suggests that the
long-range nature distorts the incident component of the relative wave packet
even at a semiclassical level.


{\it Summary}---
This study has developed a novel practical method of deducing the
induced
interactions between two impurities from a measured profile of the polaron wave packet
at several snapshots of a two-polaron collisional dynamics in the case in which impurities and medium atoms interact only weakly.
The key to success in this method is the validity of the WKB approximation, which is
satisfied for sufficiently spread polaron wave packets.
We have successfully demonstrated by solving the TDSE the possibility of reproducing such 
interactions as the RKKY, Yukawa, and Efimov type from the measured profile of the wave packets via the quantum potential.

In real possible experiments, one would be prepared to confine a degenerate
Fermi gas or Bose condensate with a sufficiently large trap and treat two non-interacting Bose gases with different internal states as impurities by putting each of them in a sufficiently narrow trap that is located at a symmetrical position with respect to the large trap of a majority gas and then by releasing the two Bose gases from the trap at the same time.
We expect that observation of transmitted wave packets after the collision would be possible
owing to the different internal states, which would help to distinguish between two impurities.
We emphasize that our strategy to deduce the mediated interaction from the probability density of the two-polaron wavefunction is different from the method of estimating the potential from a phase shift as used in an inverse scattering method \cite{InverseScattering}.

In general, the full profile of the medium-induced non-local interaction that occurs between particles is theoretically unknown.
In this regard, we believe that the method proposed here could help to easily and quantitatively obtain information
regarding mediated interactions from experiments.
It is nevertheless foreseeable that in actual experiments the interaction cannot be described by the type of mediated interactions
used in this study due to various effects such as the finite volume one.
If the experiment is carried out under appropriate initial conditions, however, the resultant deviation between the empirically
deduced interactions and the full interaction could be duly reduced.
Also, our proposed method might open an opportunity to study
nuclear interactions that are in principle microscopically
known from QCD \cite{Ishii2007, Ishii2012, Iritani2019}, from a different viewpoint that utilizes a cold atomic system as a quantum simulator.
As a next step, we will attempt to test our method for 3D collision systems and investigate the applicability to general interactions.

{\it Acknowledgments}---
We would like to thank K.~Nishimura, T.~Hata, K.~Ochi, and Y.~Yamanaka for useful discussions.
This work was supported in part by Grants-in-Aid for Scientific Research from JSPS (Nos.\
 17K05445, 18K03501, 18H05406, 18H01211, and 19K14619).


\newpage
{\bf Supplemental online material for:
``Extracting non-local inter-polaron interactions from collisional dynamics"}
\\
\par
In the supplimental material, we show the list of movies.
\\
\par
List of movies:
\begin{itemize}
  \item[1.] File: \verb#RKKY_x0_050_eta_010.mp4#

  Dynamics of the relative density and minus the quantum potential that corresponds to Figs.~2(a) and 3(a).
  \\
  Youtube: \verb#https://youtu.be/By_ZALuI6A8#

  \item[2.] File: \verb#RKKY_x0_050_eta_005.mp4#

  Dynamics of the relative density and minus the quantum potential that corresponds to Figs.~2(b) and 3(b).
  \\
  Youtube: \verb#https://youtu.be/dTKHjs05SfQ#

  \item[3.] File: \verb#RKKY_x0_025_eta_005.mp4#

  Dynamics of the relative density and minus the quantum potential that corresponds to Figs.~2(c) and 3(c).
  \\
  Youtube: \verb#https://youtu.be/9iXiWWM9H6o#

  \item[4.] File: \verb#RKKY_x0_025_eta_001.mp4#

  Dynamics of the relative density and minus the quantum potential that corresponds to Figs.~2(d) and 3(d).
  \\
  Youtube: \verb#https://youtu.be/Q74kZCrYWrY#

  \item[5.] File: \verb#Yukawa_x0_025_eta_005.mp4#

  Dynamics of the relative density and minus the quantum potential that corresponds to Fig.~4(a).
  \\
  Youtube: \verb#https://youtu.be/nTRTT_8gGME#

  \item[6.] File: \verb#Yukawa_x0_100_eta_005.mp4#

  Dynamics of the relative density and minus the quantum potential that corresponds to Fig.~4(b).
  \\
  Youtube: \verb#https://youtu.be/tJkNVh91SrY#

  \item[7.] File: \verb#Efimov_x0_025_eta_005.mp4#

  Dynamics of the relative density and minus the quantum potential that corresponds to Fig.~4(c).
  \\
  Youtube: \verb#https://youtu.be/fK-G-StErAM#

  \item[8.] File: \verb#Efimov_x0_100_eta_005.mp4#

  Dynamics of the relative density and minus the quantum potential that corresponds to Fig.~4(d).
  \\
  Youtube: \verb#https://youtu.be/2RqUnCu-Shg#
\end{itemize}

\end{document}